\documentclass{articles_perso}

\usepackage{etoolbox}
\def\bfnabla{\mbox{\boldmath $\nabla$}}
\def\calNL{\mc{N\!L}}

\newcommand{\Rm}{{\rm Rm}}
\renewcommand{\Re}{{\rm Re\,}}
\renewcommand{\Im}{{\rm Im\,}}
\newcommand{\sign}{{\rm sign\,}}
\graphicspath{{Revisiting_the_ABC_flow_dynamo_fichiers/}}

\title{Revisiting the ABC flow dynamo}
\addauth{Emmanuel Dormy}{MAG (ENS/IPGP),
  LRA, D\'epartement de Physique,
  Ecole Normale Sup\'erieure,
  24, rue Lhomond,
  75231 Paris Cedex 05, France.}{dormy@phys.ens.fr}
\begin{document}

\begin{abstract}
  The ABC flow is a prototype for fast dynamo action, essential to the
  origin of magnetic field in large astrophysical objects. Probably the
  most studied configuration is the classical \(1\!:\!1\!:\!1\) flow. We
  investigate its dynamo properties varying the magnetic Reynolds number
  \(\Rm\). We identify two kinks in the growth rate, which correspond
  respectively to an eigenvalue crossing and to an eigenvalue
  coalescence. The dominant eigenvalue becomes purely real for a finite
  value of the control parameter. Finally we show that even for \(\Rm =
  25000\), the dominant eigenvalue has not yet reached an asymptotic
  behaviour. Its still varies very significantly with the controlling
  parameter. Even at these very large values of \(\Rm\) the fast dynamo
  property of this flow cannot yet be established.
\end{abstract}

\section{Introduction}

We investigate the kinematic dynamo action associated with the well known
ABC-flow (which stands for Arnold, Beltrami, Childress, see
\citet{dombre1986chaotic}). We focus on the highly symmetric, and most
classical setup: \(A\!:\!B\!:\!C=1\!:\!1\!:\!1\)~. Its dynamo properties
have been assessed in 1981 by \citet{arnol1981magnetic} and it represents
since then the prototype flow for fast dynamo action. A ``fast
dynamo''\cite{childress1995stretch} is a flow which achieves exponential
magnetic field amplification over a typical time related to the advective
timescale and not the ohmic diffusive timescale (in which case it is
referred to as a ``slow dynamo''). It is known\cite{vishik1989mfg} that
exponential stretching of fluid elements is necessary for fast dynamo
action. The existence of fast dynamos is essential to account for the
presence of magnetic field in astrophysical bodies, for which the ohmic
diffusive time is often larger than the age of their formation. If
self-excited dynamo action is to generate their magnetic fields, it is
therefore essential that it be achieved over an advective timescale. The
most classical flow to exemplify such ``fast dynamo'' action is indeed the
ABC-flow. \citet{arnold1983growth} first investigated the dynamo property
of the ABC-flow, originally introduced to investigate Lagrangian chaos.
Many developments followed, which will be discussed in the course of this
article\cite{galloway1984numerical,galloway1986dynamo,galloway1987note}.

Most of the recent developments in this field involve non-linear studies
with forcing belonging to the class of ABC
flows\cite{courvoisier2005dynamo,archontis2007nonlinear}, with a few
noticeable exceptions\cite{alexakis2011searching,galloway2012abc}. The
asymptotic behaviour of one of the most classical example of fast dynamo is
however still not understood. This motivates the following high-resolution
linear study.

\section{Numerical method}
We are concerned with the kinematic dynamo problem, for which a solenoidal
magnetic field evolution is governed under a prescribed flow by the
induction equation
\begin{equation}
  \frac{\partial \bf B}{\partial t} = \bfnabla \times \left( \bf u \times \bf B
    - \Rm^{-1}\, \bfnabla \times \bf B \right) \,.
  \label{induct}
\end{equation}
We consider here the ABC-flow
(\citet{arnold1965topologie,henon1966topologie}), which takes the form
\begin{equation}
  \bf u = 
  (A \sin z + C \cos y) \, \bf e_x
  + (B \sin x + A \cos z) \, \bf e_y
  + (C \sin y + B \cos x) \, \bf e_z \, ,
\end{equation} 
and restrict our attention to the case where the magnetic field has the
same periodicity as the flow (i.e. \(2\pi\)-periodic in all directions of
space, see \citet{archontis2003numerical} for extensions) and the weight of
the three symmetric Beltrami components are of equal strength
(\(A=B=C\equiv 1\)).

Let us stress again that we also restrict our attention to the kinematic
dynamo problem, in which the flow is analytically prescribed and unaltered
by the magnetic field (see \citet{galloway1987note} for an investigation of
the stability of this flow).

The choice \(A\!:\!B\!:\!C = 1\!:\!1\!:\!1\) belongs to the largest
symmetry class for this kind of flows, and has for this reason been the
most intensively studied. However, it yields very small chaotic regions and
is thus possibly non optimal for dynamo action (see
\citet{alexakis2011searching} for a detailed study of this point).

The simulations presented in this article were performed using a modified
version of a code originally developed by \citet{galloway1984numerical} and
which uses a fully spectral method with explicit mode coupling.

The original time-stepping used by \citet{galloway1984numerical} relies on
a Leapfrog scheme stabilised by a Dufort-Frankel discretization of the
diffusive term. Introducing \(\mc L\) to denote the discretized diffusion
operator, which is local in Fourier space, and \(\calNL\) to denote the
discretized inductive term, non-local as it couples neighbouring modes,
this scheme can be expressed as
\begin{equation}
  \bf B^{n+1}=\bf B^{n-1}+2{\rm d}t \, \left( \calNL(\bf B^{n}) + \frac{1}{2}
    \mc L (\bf B^{n+1}+\bf B^{n-1}) \right) \, ,
\end{equation}
using a red-black (or Chloride-Sodium) staggering in time and space, see
\citet{galloway1986dynamo}.

We have implemented two alternative time stepping schemes, in order to
assess the stability of the temporal evolution at large values of
\(\Rm\). We used a Crank-Nicholson Adams-Bashforth scheme
\begin{equation}
  \bf B^{n+1}=\bf B^{n}+{\rm d}t \, \left( \frac{1}{2} \mc L (\bf B^{n+1}+\bf B^{n}) +
    \frac{3}{2}\, \calNL(\bf B^{n})-\frac{1}{2}\, \calNL(\bf B^{n-1}) \right)\,,
\end{equation}
as well as a second order BDF discretization
\begin{equation}
  \frac{3}{2} \bf B^{n+1}=2\, \bf B^{n}-\frac{1}{2}\, \bf B^{n-1} 
  +{\rm d}t \, \left(\mc L (\bf B^{n+1}) +
    2\, \calNL(\bf B^{n})- \calNL(\bf B^{n-1}) \right)\,.
\end{equation}
These two schemes are unstaggered and involve larger memory requirements,
still offering the same complexity. All schemes are semi-implicit, but
retain an explicit marching for the non-local term in order to prevent the
resolution of a linear system at each time-step. We verified that the
results presented in this article are independent of the above choices.

The computing time obviously varies with the control parameter \(\Rm\). If
all spatial modes are computed up to a truncation \(N\), the computational
complexity scales like \(\mc O(N^4)\). Assuming the asymptotic scaling of
the magnetic field length scale, we get \(N \sim \Rm ^{1/2}\) and thus
expect a complexity growing as \(\mc O(\Rm^2)\). We have therefore derived
a parallel version of the code using the MPI library and a spectral domain
decomposition strategy to tackle larger values of \(\Rm\). This yields
shorter computing time at large resolution.

The results presented in this article were obtained with numerical
resolutions ranging from \(N=64\) for the smallest values of \(\Rm\) to
\(N=1024\) for \(\Rm=25000\). In all cases we verified that the results
reported here were unaltered by doubling the resolution. The simulations
presented here were performed on up to \(512\) cores.

It is worth stressing that the quantity \(\bfnabla \cdot \bf B \) is
obviously preserved by (\ref{induct}), and that this essential property is
retained by the discrete numerical schemes, and thus the magnetic field
remains solenoidal throughout the simulations.

We investigate a linear problem and therefore expect that, independently of
the initial conditions, the long time integration will simply reflect the
eigenmode with largest growth rate. In practice, we used two different sets
of initial conditions, either
\begin{equation}
  \bf B (t=0) \propto (\sin z - \cos y)\, 
  \bf e_x + (\sin x - \cos z)\, 
  \bf e_y + (\sin y - \cos x)\, 
  \bf e_z \, ,
\end{equation}
or a random initial condition with a spectrum converging as \({\bf
  k}^{-2}\) for regularity and projected numerically to get a non-divergent
field.

\section{Modes crossing}

This study is focused on the \(1\!:\!1\!:\!1\) ABC-flow, which has a large
number of symmetries. These have been well
documented\cite{arnold1987evolution,podvigina1999spatially}. The symmetry
group is generated by two independent rotations: the cloverleaf rotation
\(T: x\rightarrow y\rightarrow z\rightarrow x\) and a rotation of angle
\(\pi/2\) followed by a translation (\(Rt\) in the sequel) around one of
the three coordinate axes. For example \(Rt_y: x\rightarrow \pi/2+z,
y\rightarrow y-\pi/2, z\rightarrow \pi/2-x\). All the other symmetries are
obtained by combinations of these two rotations. The resulting group of
symmetries of the ABC-flow contains 24 elements (including identity).
We tested our numerical results against all the symmetries of the ABC-flow.

\begin{figure}
  \centering
  \includegraphics[width=0.8\textwidth,clip]{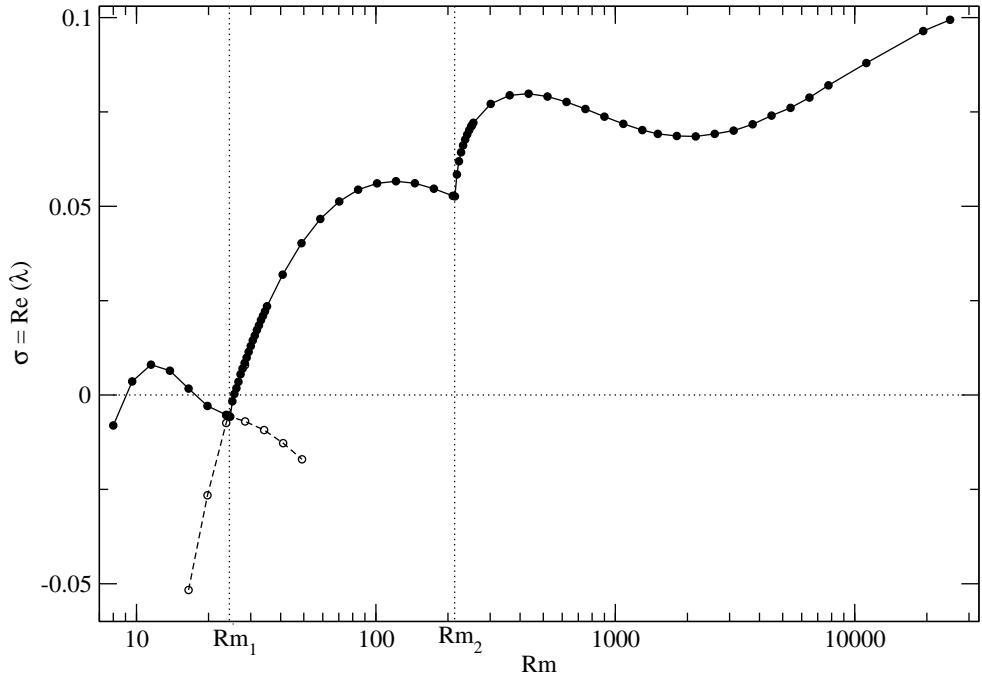}
  \caption{\label{fig:taux} Plot of the real part of the eigenvalue for the
    fastest growing magnetic field mode as a function of the magnetic
    Reynolds number \(\Rm\) (using logarithmic scale in the \(x\)-axis).}
\end{figure}

Figure~\ref{fig:taux} presents the evolution of the maximum growth rate of
the magnetic field as a function of \(\Rm\). Each point on the figure
corresponds to a three-dimensional simulation. We confirm growth rates
obtained by earlier studies (see \citet{galloway2012abc} for a recent
review), and we extend the range of investigation from \(\Rm < 1600\) to
\(\Rm < 25000\). The curve has been validated against published growth
rates using spectral methods\cite{galloway1986dynamo,lau1993fast} as well
as a finite volume method \cite{teyssier2006kinematic} for which
simulations have been performed up to \(\Rm=2000\) (Teyssier \& Dormy
private comm.).

\begin{figure}
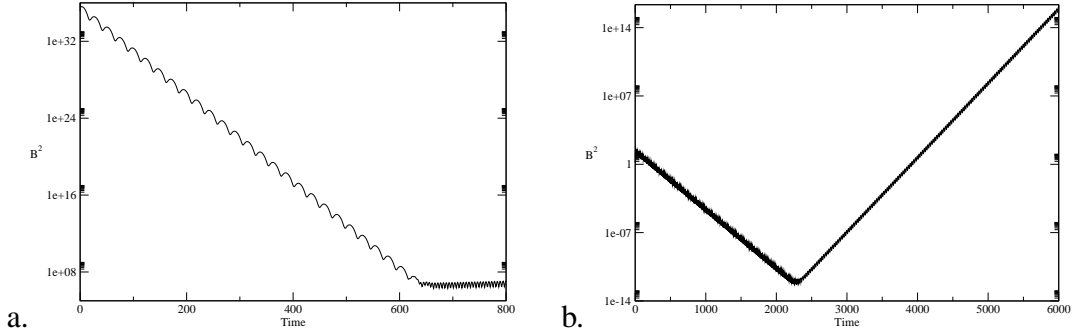

  \centering
  a.\includegraphics[width=0.4\textwidth,clip]{16_5}\ \hskip 5mm
  b.\includegraphics[width=0.4\textwidth,clip]{28_4}
  \caption{\label{fig:transit} Time evolution of the magnetic energy, which
    highlights the information provided by transient behaviours. Left (a) a
    simulation for \(\Rm = 16.5 \, (<\Rm_1)\) using as initial condition
    the final solution obtained at \(\Rm > \Rm_1\), right (b) simulation
    for \(\Rm = 28.4 \, (>\Rm_1)\).}
\end{figure}

In addition to the wider extend of \(\Rm\) variation, our curve also offers
a finer resolution than previously obtained graphs. This highlights the
presence for two kinks in the curve, labelled \(\Rm_1\) and \(\Rm_2\) on
the figure. The first of these occurs in the stable window reported by
\citet{galloway1986dynamo} near \(\Rm=20\) and corresponds to \(\Rm_{1}\in
[24.05, 24.10]\). A mode crossing was previously suggested owing to the
changes in the eigenfunction symmetry\cite{galloway2012abc}. Here we
demonstrate this eigenvalue crossing by following both eigenvalues on each
side of the crossing. In fact whereas time stepping algorithms usually only
provide information on the dominant eigenvalue, i.e.\ the eigenmode with
largest growth rate, we use it here to get more information. Indeed,
transient behaviour starting with well selected initial conditions provide
information on the behaviour of a given mode, even if it is not the
dominant eigenmode (see Figure~\ref{fig:transit}). This transient behaviour
allowed us to continue the branches corresponding to each eigenvalue
outside of the region in which they are dominant eigenvalues (see dotted
lines and open symbols on Figure~\ref{fig:taux}).

In the first window \(\Rm < \Rm_1\) as argued by
Arnol'd\cite{arnold1987evolution}, we observe that the dominant eigenmode
has all ``even'' symmetries of the ABC-flow, i.e.\ it has every combination
of an even number of \(Rt\) as a symmetry, and is antisymmetric otherwise
(the solution changes sign by the corresponding transformation). In the
second window \(\Rm > \Rm_1\), we observe numerically that all the above
symmetries and anti-symmetries disappear, as pointed by
\citet{galloway1986dynamo}. However, \citet{JonesGilbert} are currently
using a decomposition of this mode in three components each satisfying
different symmetries.

\begin{figure}
  \centering
  \includegraphics[width=0.8\textwidth,clip]{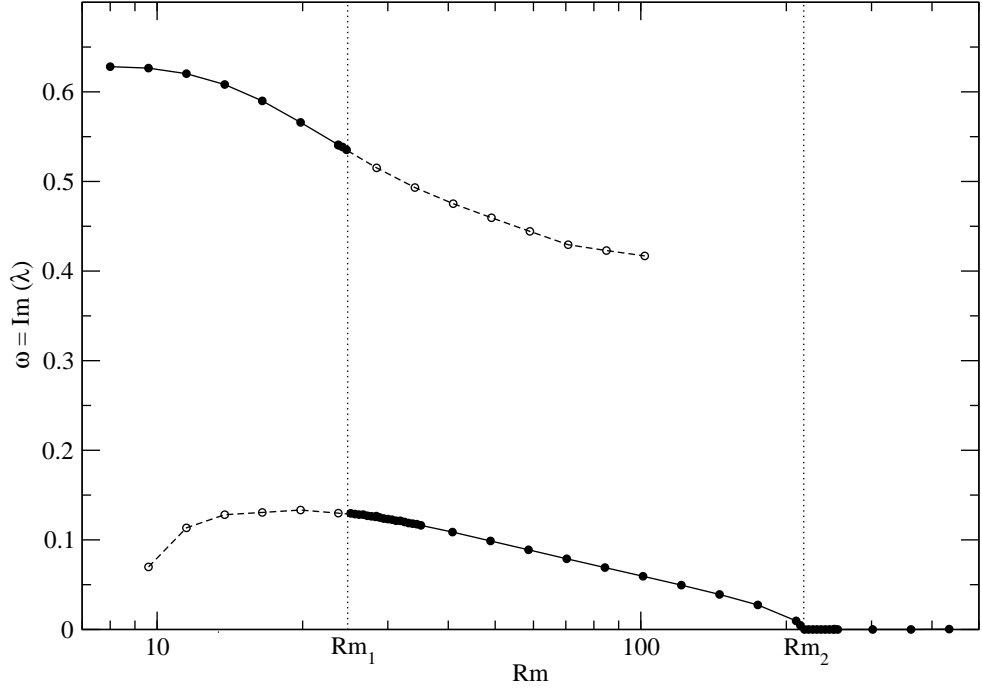}
  \caption{\label{fig:oscs} Plot of the imaginary part of the eigenvalue as
    a function of Rm (using logarithmic scale in the \(x\)-axis).}
\end{figure}

As noted by earlier authors, the dominant eigenvalues are complex, leading
to oscillations of the energy, visible on Figure~\ref{fig:transit}
(particularly on the first part of Figure~\ref{fig:transit}a, as the period
of oscillations is elsewhere very short compared to the time extend of the
plot). The imaginary part of the dominant eigenvalue \(\omega=\Im
(\lambda)\) can thus be directly determined from these time series. The
graph \(\omega(\Rm)\) is displayed on Figure~\ref{fig:oscs}. As explained
above, not only do we display the dominant eigenvalue (solid line and
symbols) but we are also able to follow each mode past their region of
selection (dotted lines and open symbols). One can note that, as \(\Rm\)
increases past \(\Rm_1\), the imaginary part of the dominant eigenvalue
jumps discontinuously from \(\omega \simeq 0.53\), corresponding to the
first window identified by \citet{arnold1983growth}, to \(\omega \simeq
0.13\) corresponding to the second window of
\citet{galloway1986dynamo}. This discontinuous jump in the pulsation
highlights the eigenvalue crossing occurring at \(\Rm_1\). There again,
transient behaviours were used to obtain the open symbols.

\section{Oscillatory dynamics}
In order to improve our understanding of the oscillatory dynamics, we
introduce a phase space for this linear system. We rely for \(\Rm < \Rm_1\)
on two vectors, corresponding respectively to the dominant \(|{\bf k}|=1\)
contribution\cite{arnold1987evolution}
\begin{align}
  \bf b_1 &=(\sin(z) - \cos(y))\, \bf e_x + (\sin(x) - \cos(z))\, \bf e_y
  + (\sin(y) - \cos(x))\, \bf e_z \, ,\\
\intertext{and to \(|{\bf k}|=2\)}
  \bf b_2 &=(\sin(y)\cos(z))\, \bf e_x + (\sin(z)\cos(x))\, \bf e_y +
  (\sin(x)\cos(y))\, \bf e_z \,.
\end{align} 
These two vector fields satisfy all symmetries of the realised eigenmode
for this parameter regime (\(\Rm < \Rm_1\)).\\
Simulations in this regime rapidly reach an asymptotic behaviour starting
with \({\bf b_1}\) as an initial condition.

\begin{figure}
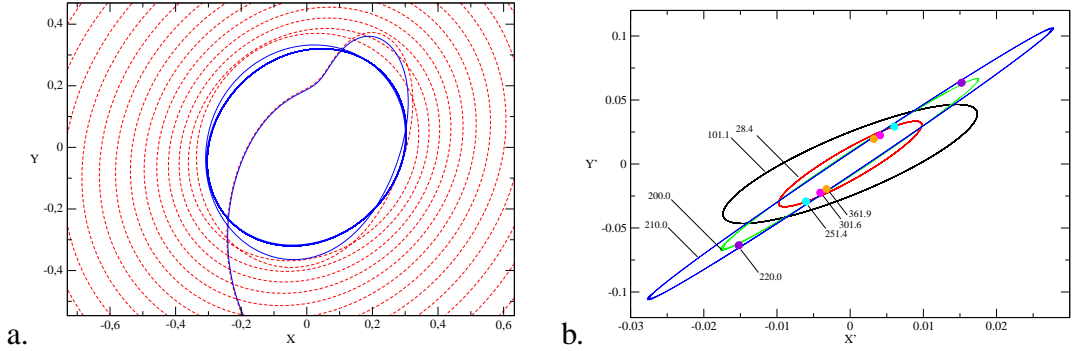

  \centering
  a.\includegraphics[width=0.4\textwidth,clip]{ellipse_11_5}
  \hskip 5mm
  b.\includegraphics[width=0.4\textwidth,clip]{ellipse_2e_fenetre_tronc}

  \caption{\label{fig:ellipse} Phase space diagrams for \(\Rm < \Rm_1\) (a)
    and \(\Rm > \Rm_1\) (b).}
\end{figure}

We construct the phases by introducing
\begin{equation}
  X=\langle{\bf B}\cdot {\bf b}_1\rangle
  \, \exp^{-\sigma t}\, ,
  \quad \text{and} \quad Y=\langle{\bf B}\cdot {\bf b}_2\rangle
  \, \exp^{-\sigma t}\, ,
  \label{eq:projection}
\end{equation} 
where the growth rate \(\sigma\) is a function of \(\Rm\) (see
figure~\ref{fig:taux}). As the governing equations are linear, the
exponential damping is here essential in order to introduce a limit
behaviour. The quantities \(X\) and \(Y\) are presented on
figure~\ref{fig:ellipse}a for \(\Rm=11.5\). The undamped trajectory of the
system is also represented using dashed lines, and directly illustrates the
exponential growth of the dominant mode. The damped trajectory evolves
toward the equivalent of a stable limit cycle.

The exponential damping on \(X\) and \(Y\) provides us with the equivalent
of a non-linear dynamical system. This explains why we report below
phenomena that are usually associated to non-linear dynamics.

In the second window, \(\Rm > \Rm_1\), the two modes we selected
(corresponding to the lowest \(\vec k\) component of the realised mode) are
\begin{align}
  \bf b'_1 &= (-\sin(y) - \cos(z)) \, \bf e_x + (\sin(z) + \cos(x))\, \bf e_y
  + (-\sin(x) + \cos(y))\, \bf e_z \, ,\\
  \bf b'_2 &=( \sin(y) - \cos(z)) \, \bf e_x + (\sin(z) - \cos(x))\, \bf e_y
  + (\sin(x) - \cos(y))\, \bf e_z \,.
\end{align} 
These correspond to two components of the general family reported earlier
for this mode\cite{galloway1986dynamo}. These two modes involve symmetries
(though none in common), but these are not relevant here since these are
not verified by the full eigenmode.

The resulting orbits of \(X'=\langle{\bf B}\cdot {\bf b'}_1\rangle \,
\exp^{-\sigma t}\) and \(Y'=\langle{\bf B}\cdot {\bf b'}_2\rangle \,
\exp^{-\sigma t}\) are represented on figure~\ref{fig:ellipse}b for various
values of the controlling parameter \(\Rm>\Rm_1\). The represented
quantities are not arbitrarily rescaled. Only the exponential damping has
been applied, and all cases have been started with the same initial
condition (involving random, but divergence free, fluctuations). The
oscillating nature of the dynamo for \(\Rm < \Rm_2\) is clearly illustrated
by the limit cycle. For \(\Rm > \Rm_2\) the oscillations disappear and the
dynamo mode therefore becomes a fixed point in the \([X', Y']\) plan. For
clarity, we suppressed the trajectories that lead to the limit cycles or
the steady solutions on figure~\ref{fig:ellipse}b.

\section{Eigenvalues coalescence}
As the magnetic Reynolds number is further increased, a second kink in the
growth rate is observed on Figure~\ref{fig:taux} for \(\Rm_{2}\in [215.0,
215.4]\). This second accident, however, does not correspond to a change of
dominant eigenvalue, but instead to an eigenvalues coalescence. The
strategy highlighted above to follow secondary modes is inefficient here,
indicating that there is no significant change in the dominant eigenmode.

Figure~\ref{fig:oscs} reveals that the behaviour of the imaginary part of
the eigenvalue is very different near the second kink. Instead of the
abrupt jump reported at \(\Rm_1\), the pulsation continuously (but not
smoothly) tends to zero as \(\Rm\) approaches \(\Rm_2\) and vanishes for
\(\Rm > \Rm_2\).

The lack of oscillations at large \(\Rm\) is a well known characteristic,
it was already noticed by \citet{galloway1984numerical} for (\(\Rm >400\)),
although they could not assess whether the period of oscillations was
simply increasing with \(\Rm\) or the eigenvalue had become purely real.
\citet{lau1993fast} suggested that this could be associated with a mode
crossing, a new mode with purely real eigenvalue taking over above
\(\Rm_2\).

We show here that the imaginary part of the eigenvalue indeed vanishes for
\(\Rm > \Rm_2\), and that this corresponds to the coalescence of two
complex conjugate eigenvalues on the real axis. The coalescence yields the
kink in the evolution of the real part of the eigenvalue.

The simplest mathematical model for a complex conjugate eigenvalue
coalescence on the real axis corresponds to a situation of the form
\begin{equation}
  \lambda_\pm = \alpha(\Rm) \pm \sqrt{\beta(\Rm)}\, ,
  \label{coalesc}
\end{equation}
where \(\alpha\) and \(\beta\) are differentiable real functions of
\(\Rm\). A negative \(\beta\) (for \(\Rm < \Rm_{2}\)) yields two complex
conjugate modes, and thus oscillations of the magnetic energy. As \(\beta\)
becomes positive (for \(\Rm > \Rm_{2}\)), the eigenvalues are purely real
and the \(\beta\) term now contributes to the real part of the eigenvalue
\(\lambda_+\) offering the largest growth rate.

\begin{figure}
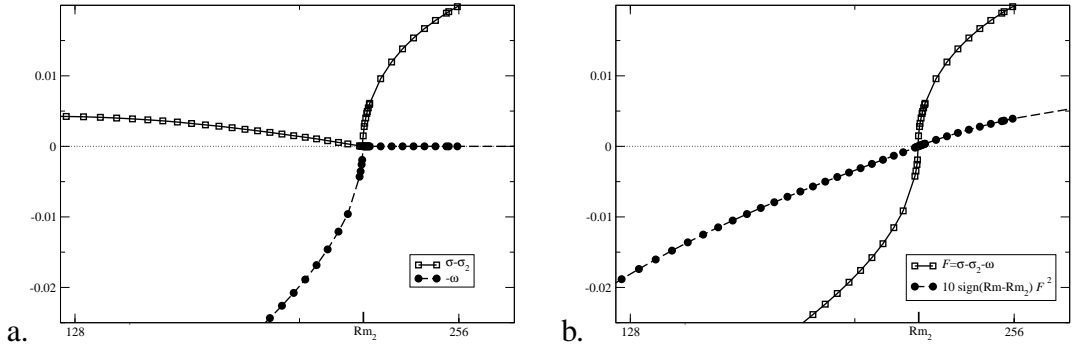

  \centering
  a.\includegraphics[width=0.4\textwidth,clip]{abc_croisement}\ \hskip 5mm
  b.\includegraphics[width=0.4\textwidth,clip]{abc_croisement2}
  \caption{\label{fig:croisement} Eigenvalues coalescence for \(\Rm\) close
    to \(\Rm_2 \simeq 215\). Both the real part (written as
    \(\sigma(\Rm)-\sigma_2\)) and the opposite of the imaginary part
    (i.e. \(-\omega(\Rm)\)) are represented (a). The sum \(F\) of both (b),
    and \(\sign(\Rm-\Rm_2)F^2\) illustrate the continuity and regularity in
    the functional form of the eigenvalue (see text).}
\end{figure}

Figure~\ref{fig:croisement}a presents a detailed view on the variation of
\(\sigma = \Re(\lambda)\) and \(\omega = \Im(\lambda)\) close to \(\Rm_2\).
Defining \(\sigma_2=\sigma(\Rm_2)\) we plot \(\sigma(\Rm)-\sigma_2\) and
\(-\omega(\Rm)\). It is clear that the kink in \(\sigma\) is concomitant of
the vanishing of \(\omega\).

Let us now form on Figure~\ref{fig:croisement}b the quantity \(F =
\sigma(\Rm)-\sigma_2-\omega(\Rm)\). The square-root behaviour of \(F\) near
\(\Rm_2\) is obvious. Assuming that the above model (\ref{coalesc}) is
correct, \(F\) corresponds to
\(\alpha-\alpha_0+\sign(\beta)\sqrt{|\beta|}\), where
\(\alpha_0=\alpha(\Rm_2)\). We can note on Figure~\ref{fig:croisement}a for
\(\Rm < \Rm_{2}\) that \(\alpha-\alpha_0\) remains small compared to
variations in \(\beta\). The quantity \(\sign (\Rm-\Rm_2) \, F^2\)
therefore offers a good approximation to \(\beta\) and should be
differentiable at \(\Rm_2\).

More formally, assuming that \(\alpha\) and \(\beta\) are regular functions
of \(\Rm\), we can write a finite expansion of the form
\begin{align}
  \alpha &= \alpha_0 + \alpha_1(\Rm-\Rm_2) + \alpha_2(\Rm-\Rm_2)^2+\cdots,\\
  \beta &=\beta_1(\Rm-\Rm_2) + \beta_2(\Rm-\Rm_2)^2+\cdots,
\end{align}
with \(\beta_1>0\). The quantity \(\sign(\Rm-\Rm_2)\,F^2\) can be written
at the lowest orders in \(\Rm-\Rm_2\)
\begin{equation}
  \begin{split}
    \sign(\Rm-\Rm_2)\,F^2 &= \sign(\Rm-\Rm_2)\alpha_1^2(\Rm-\Rm_2)^2 \\
    &+ \sign(\Rm-\Rm_2)\alpha_1\sqrt\beta_1(|\Rm-\Rm_2|)^{3/2} \\
    &+ \beta_1(\Rm-\Rm_2).
  \end{split}
\end{equation}
This development implies that \(\sign(\Rm-\Rm_2) \, F^2\) is differentiable
at \(\Rm_2\). Figure~\ref{fig:croisement}b clearly illustrate this property
on the direct numerical simulation.

Another insight on the nature of this transition can be gained from the
``phase space'' introduced in figure~\ref{fig:ellipse}b. As noted above a
remarkable feature is that the steady solutions obtained after the
coalescence lies on the ellipse described by the limit cycle shortly below
the coalescence (on the figure \(\Rm = 220\)). This behaviour which is
similar to that of an excitable system (such as a pendulum subject to a
constant torque), has recently been observed in experimental dynamos
measurements\cite{ravelet2008chaotic}. The phase space was then constructed
using two components of the magnetic field at a given location (a probe).

\begin{figure}
  \centering
  \includegraphics[width=0.6\textwidth,clip]{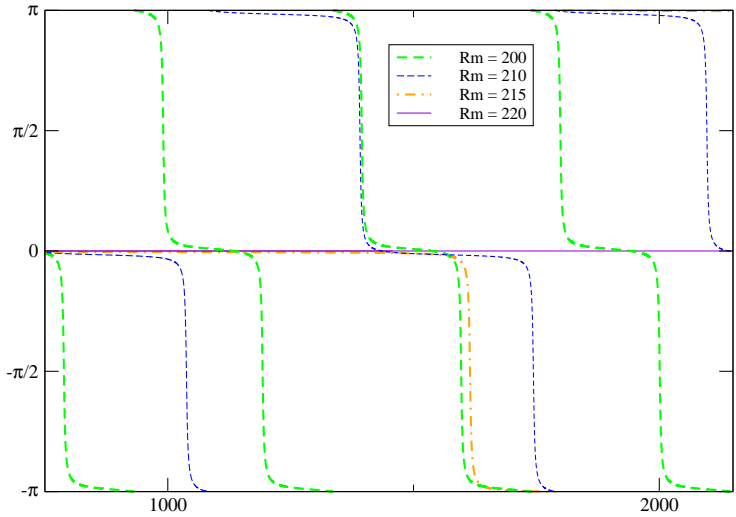}
  \caption{\label{fig:theta} Evolution in time of the projection of the
    dominant mode on the two large scale components, represented via the
    angle \(\theta\), such that \((\cos (\theta),\sin(\theta)) = (X',Y')/\sqrt{X'^2+Y'^2}\).}
\end{figure}

Such behaviour is reminiscent of a saddle-node bifurcation. In these
systems, the dynamics becomes increasingly slow on the cycle as the system
approaches the state at which the saddle and the node will collapse. In
order to assess this property in our system, albeit linear, we introduce
the angle \(\theta\) of the system over a unit circle described through the
orbit, so that \((\cos (\theta),\sin(\theta)) = (X',Y')/\sqrt{X'^2+Y'^2}\).
The time evolution of \(\theta\) with increasing values of \(\Rm\) is
presented in figure~\ref{fig:theta}. The system clearly spends an
increasing amount of time as \(\Rm\) approaches \(\Rm_2\) near the angle at
which the stable solution will occur for \(\Rm=\Rm_2\). Such behaviour
could be described by a simple phase dynamics,
e.g. \citet{petrelis2009simple,guckenheimer1997nonlinear}.

It is interesting that these approaches of non-linear dynamics can cast
some light on the behaviour of kinematic dynamos. A similar occurrence of a
saddle-node transition for a kinematic dynamo numerical model of the above
mentioned VKS experiment has indeed been reported in
\citet{gissinger2009numerical}, with an expression of the form
\eqref{coalesc}.

\begin{figure}
  \centering
  \includegraphics[width=0.4\textwidth,clip]{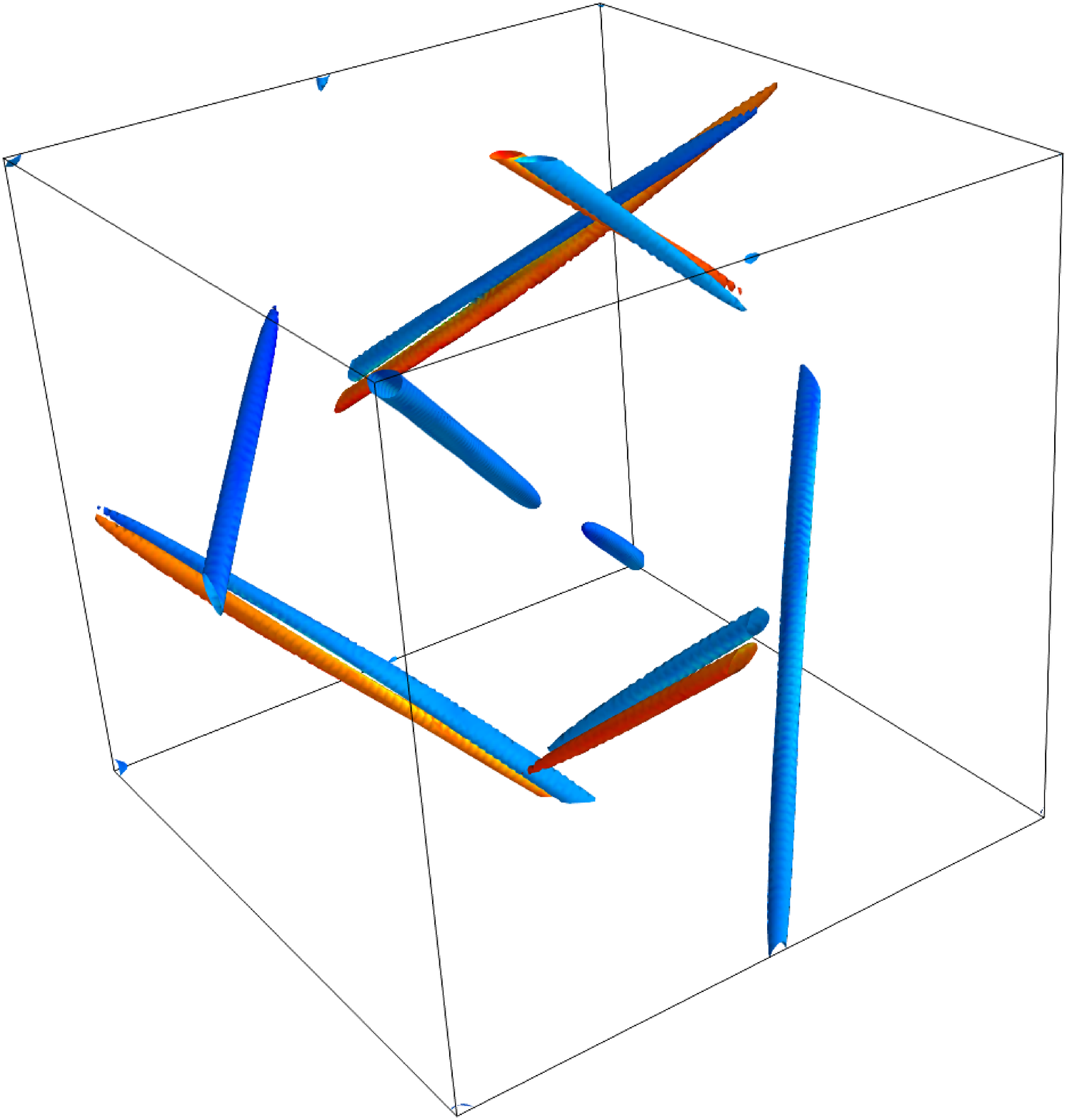}\qquad
  \includegraphics[width=0.4\textwidth,clip]{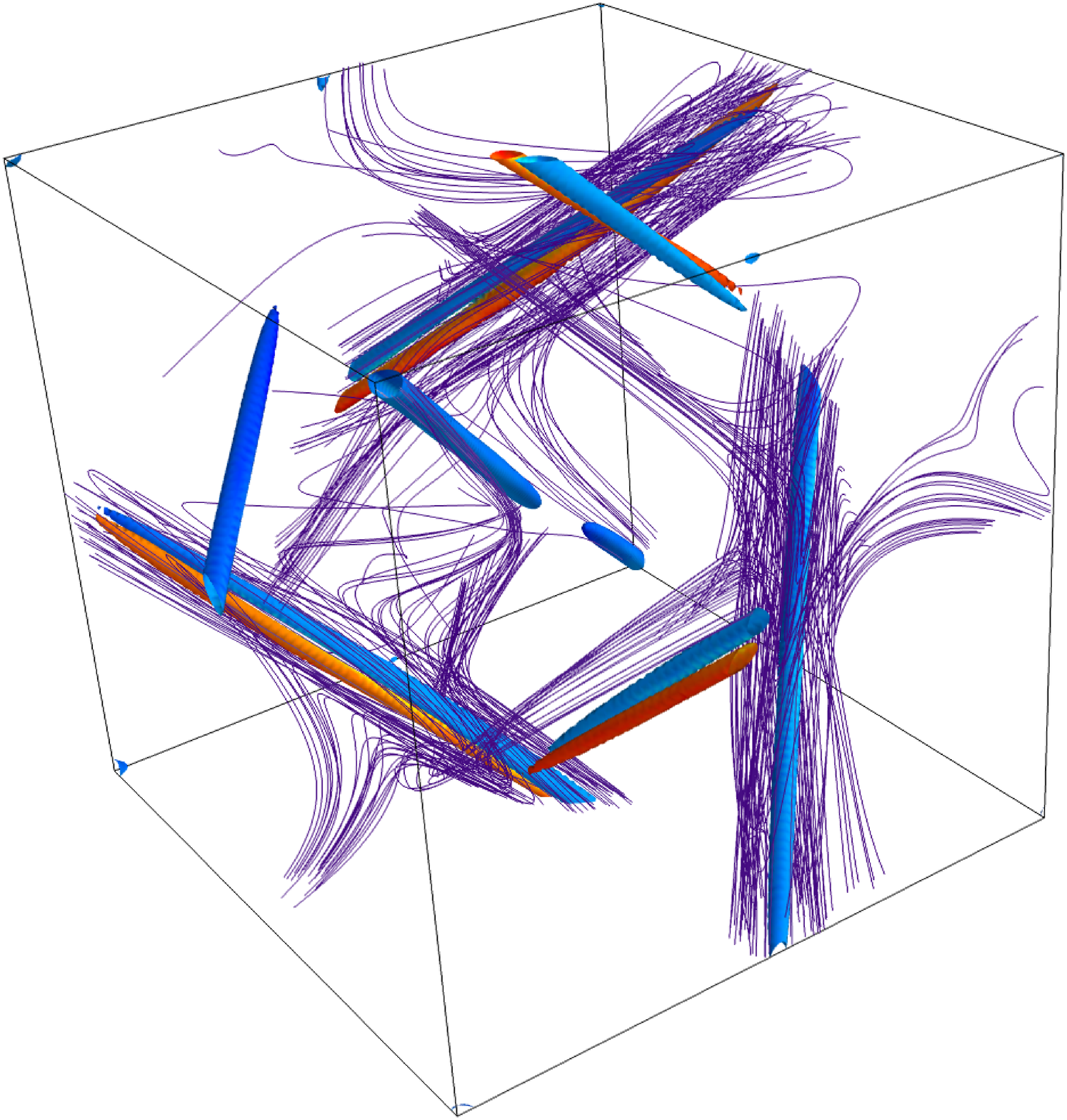}
  \caption{\label{fig:croisement3D} Eigenmode obtained for \(\Rm=434.2 \,
    (>\Rm_2)\). An isosurface of the magnetic energy is represented. To
    highlight the symmetry, it is coloured according to the value of
    \(B_x\), from blue (negative) to red (positive). The corresponding
    eigenvalue is purely real, yet the double cigars structure of the field
    remains clearly visible. Magnetic field lines are also depicted on the
    right plot.}
\end{figure}

The fact that the fixed points in the \((X',Y')\) plan describe the
previously existing limit cycle is a strong indication that there is at
first no significant change occurring in the structure of the dominant
eigenmode after the eigenvalues coalescence. Indeed the ``double cigars''
structure (see \citet{dorch2000structure}), associated to the oscillations
for \(\Rm \in [\Rm_1, \Rm_2 ]\) is preserved once the growth rate has
become steady, \( \Rm > \Rm_2\) (see Figure~\ref{fig:croisement3D}).

\section{Asymptotic behaviour}
We have finally increased the control parameter in the range
\(215\)--\(25000\). Despite the fact that the largest magnetic Reynolds
number tackled in this study is roughly 15 times larger than earlier
results, the growth rate has not reached an asymptotic value yet. The
growth rate obtained for our largest \(\Rm\) is very close to \(0.1\) and
appears to be still significantly increasing with \(\Rm\).

The \(1\!:\!1\!:\!1\) ABC-flow has also been considered by
\citet{gilbert1992magnetic} using maps in a limit in which the diffusivity
is formally set to zero. This approach has yield growth rate of \(0.04\) --
\(0.05\), so much smaller than the value achieved by our direct numerical
simulations at \(\Rm = 25000\). It is therefore not unplausible to
anticipate that the behaviour of \(\sigma(\Rm)\) above \(25000\) will not
be monotonic and \(\sigma\) will probably decrease again.

Another indication is provided by the largest Lyapunov exponent of the
flow, which is approximately \(0.055\) (see \citet{galanti1992linear}).
Owing to the lack of regularity of the field in the limit of large \(\Rm\)
numbers, the largest Lyapunov exponent however does not provide an upper
bound on the asymptotic growth rate\cite{childress1995stretch}. An upper
bound can be sought by considering the topological entropy \(h_{\rm top}\)
(see \citet{finn1988chaotic,finn1988chaotic2}). For steady
three-dimensional flows the topological entropy is equal to the line
stretching exponent \(h_{\rm line}\) (see \citet{childress1995stretch}),
which can be estimated for the \(1\!:\!1\!:\!1\) ABC-flow to be \(h_{\rm
  line} \simeq 0.09\). This provides yet another indication that the curve
\(\sigma(\Rm)\) must decrease for larger values of \(\Rm\).

A plausible scenario, suggested by the behaviour of submodes as
investigated by \citet{JonesGilbert}, is that two complex conjugate
eigenvalues may emerge again at larger \(\Rm\). This is often observed in
saddle-node bifurcations (e.g. \citet{ravelet2008chaotic}). It would result
in the reappearance of the oscillations, and an abrupt decrease of the
growth rate (the counterpart of the increase observed at \(\Rm_2\)). This
would deserve further study.

The asymptotic behaviour of the \(1\!:\!1\!:\!1\) ABC-flow is thus not yet
established. It is at the moment, despite the high resolution simulations
presented here, impossible to assess its asymptotic growth rate. It is not
even possible to rule out the possibility of an eventual decay of the
growth rate to zero at very large \(\Rm\).

\section{Conclusion}

We have investigated using high resolution direct numerical simulations the
behaviour of the \(1\!:\!1\!:\!1\) ABC-dynamo. We have shown that the two
dynamo windows identified for this dynamo are associated with a change of
dominant eigenvalue. We have identified a second kink in the growth rate as
a function of \(\Rm\) and shown that it corresponds to an eigenvalue
coalescence and the end of the oscillatory nature of the solutions.
Finally, even at very large values of \(\Rm\), we show that the growth rate
is still strongly varying and not monotonic yet.

Relaxing the requirement of a fully three-dimensional flow and allowing for
time dependence, other models for fast dynamo actions have been obtained by
\citet{galloway1992numerical}, with a velocity depending only on two
coordinates. The time dependence ensures exponential stretching at least in
this plane. The induction equation is then separable in the $z$ direction,
allowing faster numerical integrations. The asymptotic limit of large $\Rm$
appears easier to reach for such flows. Other studies involved time
dependence of the flow\cite{otani1993fast} and some hint at a possible
resonance phenomenon\cite{dormy2008time}.

Finding a good example of fully three-dimensional flow that acts as a fast
dynamo remains a challenging problem. The most classically given example
remains the $1\!:\!1\!:\!1$ ABC-flow. This unexpectedly rich behaviour of
the \(1\!:\!1\!:\!1\) ABC-dynamo at very large \(\Rm\), highlighted in our
study, deserves further investigations.  It is most likely associated with
the fact this flow yields very small chaotic regions\cite{galloway2012abc}.

\section*{Acknowledgements}
The authors are very grateful to Dave Galloway for sharing his original
dynamo code, which served as the starting point for the parallel version
written for this study. We are also very grateful to Christophe Gissinger
for useful discussions in the course of this work.\\
Computations were performed on the MesoPSL cluster as well as on the Cines
computing centre (Genci project LRA0633).

\AtEndEnvironment{thebibliography}{%
\bibitem[Jones et al.(2012)S.E. Jones and A.D. Gilbert]{JonesGilbert}
  S.E. Jones and A.D. Gilbert, ``Utilising symmetries in kinematic dynamo
  simulations'', presentation at the European GDR Dynamo meeting, Nice
  2012, to be published in ``Dynamo action in the ABC flows using
  symmetries''.
}
\bibliographystyle{abbrvnat}
\bibliography{biblio_perso}

\end{document}